\documentclass{mn2e}
\input{psfig.sty}

\begin{document}

\title[The lack of variability of the iron line in MCG--6-30-15:
GR effects]
{The lack of variability of the iron line in MCG--6-30-15:
general relativistic effects }
\author[G. Miniutti, A.C. Fabian, R. Goyder and A.N. Lasenby]
{G. Miniutti$^1$\thanks{E-mail:
miniutti@ast.cam.ac.uk}, A.C. Fabian$^{1}$, R. Goyder$^{2}$ and
  A.N. Lasenby$^{2}$ \\ $^1$ Institute of Astronomy, University of
  Cambridge, Madingley Road, Cambridge CB3 0HA \\ 
$^2$ Astrophysics Group, Cavendish Laboratory, Madingley Road,
Cambridge, CB3 0HE
}

\pagerange{\pageref{firstpage}--\pageref{lastpage}} \pubyear{2003}

\maketitle

\label{firstpage}

\begin{abstract}
{ The spectrum and variability of the Seyfert galaxy MCG--6-30-15 can
  be decomposed into two apparently disconnected components: a highly
  variable power law and an almost constant component which contains a
  broad and strong iron line. We explore a possible explanation of the
  puzzling lack of variability of the iron line, by assuming that the
  variations of the power law component are due to changes in the
  height of the primary source in the near vicinity of a rotating
  black hole. Due to the bending of light in the strong field of the
  central black hole, the apparent brightness of the power-law
  component can vary by about a factor $4$ according to its position,
  while the total iron line flux variability is less than $20$~per
  cent. This behaviour is obtained if the primary source is 
  located within 3--4 gravitational radii ($r_{\rm g}$) from the
  rotation axis with a variable height of between $\sim$~3 and 
  8 $r_{\rm g}$. These results revive the possibility that future X--ray
  observations of MCG--6-30-15 can map out the strong gravity regime
  of accreting black holes.}

\end{abstract}


\section{Introduction} 

The X--ray spectrum of the Seyfert 1 galaxy MCG--6-30-15 exhibits a
broad emission feature peaked at $6.4$ keV which extends from below
$4$ keV to approximately $7$ keV, and was first resolved with ASCA
\cite{tetal95}. The profile of this feature is consistent with that
expected from iron fluorescence from the surface of an accretion disc
surrounding a massive black hole.

The X--ray continuum emission of MCG--6-30-15 is highly variable
(see e.g. Vaughan, Fabian \& Nandra 2003; Fabian \& Vaughan 2003).
Since the reflection features, such as the observed iron line, are
expected to be due to reprocessing of the hard X--ray continuum by
dense gas in the accretion disc \cite{gf91,mpp91}, the line flux
should respond to the changes in the illuminating continuum (Reynolds
et al 1999). Line flux variations have been reported on timescales of
about $10^4$~s, but with a much smaller amplitude than the
observed continuum changes
\cite{ietal96,ve01,sif02,fv03}. Furthermore, the variability of the
iron line flux and the observed continuum are uncorrelated.

The source spectrum and variability can be explained by a two
component model \cite{sif02} consisting of a variable power-law
component and an almost constant reflection component containing the
iron line. This model successfully accounts for the behaviour of
MCG--6-30-15 during a $320$~ks long {\it{XMM--Newton}} observation
\cite{fv03}. Hereafter, we shall refer to these components as the
Power Law Component (PLC) and the Reflection--Dominated Component
(RDC). The existence of a varying soft component with a spectral shape
uncorrelated with flux together with an almost constant harder
component is also evident in the linearity of the
flux--flux relationship recently presented by Taylor, Uttley \&
M$^{\rm c}$Hardy (2003). 

The extreme red wing of the iron line indicates that the inner radius
of the disc extends down to only a few gravitational radii 
(Wilms et al 2001; Fabian et al 2002). Strong
gravitational field effects dominate the behaviour of photons in this
region with not only gravitational redshift being severe but also
gravitational light bending \cite{mm96,mkm00,dl01,mmk02}. Fabian \& Vaughan
(2003) have recently suggested that changes in the position of the
power--law emission region could lead to large variations in the
observed flux together with a strong and almost constant reflection
component, containing the broad iron line. This behaviour would be due
predominantly to strong light bending, which is expected if the
primary source is located close to the central black hole and
illuminates the inner regions of the accretion disc, as the shape of
the iron line suggests. As the source approaches these regions of the
disc close to the black hole so more radiation is bent away from our
line of sight and intercepted by the disc. Such light bending also
explains the high observed equivalent width of the line.

In this Letter, we investigate, for the parameters relevant to
MCG--6-30-15, the variability of the PLC and of the RDC as a
function of the position and state of motion of the primary X--ray
source, seeking solutions that allow for a strongly varying PLC and an
almost constant RDC. The main purpose of this work is to provide an
astrophysical context in which the two component model for the
variability of  MCG--6-30-15 can be explained self--consistently. 
Our calculations are performed using a
combination of Monte Carlo and ray-tracing methods that take into
account the effects of Doppler/gravitational redshifts and light
beaming/bending on the motion of photons in the spacetime of a
rotating black hole.

\section{Model and assumptions}

The observed broad iron line profile and its extended red wing 
\cite{wetal01,fetal02} imply that most of the emission 
originates at radii less than $6 r_g$ ($r_g =GM/c^2$), suggesting  that 
the central black hole is rapidly spinning
\cite{ietal96,detal97}. Our computations are then performed in the
spacetime of a maximally rotating (Kerr) black hole. 

The accretion disc is assumed to be thin and to lie in the equatorial plane,
perpendicular to the rotation axis. The accreting material is flowing
along stable circular geodesics and the disc extends from the radius
$r_{\rm{ms}}$ of the marginal stable circular orbit \cite{nt73} 
out to $r_{\rm{out}} = 100~r_g$.     

The source of primary hard X--rays is assumed to have a ring--like
axisymmetric geometry and is located above the disc at a 
distance $\rho_s$ from the rotation axis and at height $h_s$, 
related to the Boyer--Lindquist coordinates $r_s$ and $\theta_s$ of the source by
$r_s = \sqrt{\rho_s^2 + h_s^2}$ and $\theta_s = \arctan (\rho_s /
h_s)$. The source can be both static or corotating with the disc. 
If the source is static, its $4$--velocity has to be proportional to
the time--like Killing vector of the spacetime so that
\begin{equation}
{\mathbf u} = C_{\rm{stat}}\,\partial_t \ ,
\end{equation}
while if it is corotating one has
\begin{equation}
{\mathbf u} = C_{\rm{corot}}\,(\partial_t + \Omega \partial_\phi ) \ ,
\end{equation}
where $\Omega \equiv 1/[a + (r\sin\theta)^{3/2}]$ and where 
$C_{\rm{stat}}$ and $C_{\rm{corot}}$
are found by requiring that the source follows a time--like world line 
(${\mathbf u}\cdot {\mathbf u} = -1$). 

The case of a static source is
investigated mainly for comparison with the corotating one that we
consider physically more plausible. This is because since flares are
believed to be associated with magnetic activity of the disc, they are
more likely to be corotating with the accreting material rather than
static. In the case of a corotating source, since the orbital
timescale in the vicinity of the central massive black hole is much
shorter than the integration time needed in observations, any
information on the azimuthal position of the flare is lost. This
justifies our choice of a ring--like axisymmetric source rather than a
point--like whose azimuthal position would affect the iron line profile
emitted from the disc but is unobservable with current long observations.

The source produces isotropic emission in its proper frame with a
power law luminosity $L_s = L_0 E_s^{-\alpha}$, where $E_s$ is the
photon energy in the source comoving frame and the spectral index is
chosen to be $\alpha = 1.1$, consistent with the observed photon index
in the spectrum of MCG--6-30-15 (see e.g. Fabian \& Vaughan 2003). 
Isotropic emission is enforced by using the 
{\small{HEALP}}ix\footnote{see http://www.eso.org/science/healpix}
(Hierarchical Equal Area isoLatitude Pixelisation) package which
allows one to define equal area curvilinear quadrilaterals on the
sphere. Isotropy (in the source proper frame) is obtained 
by sending photons along the directions defined by the geometrical 
centre of each quadrilateral.  

The photons emitted by the primary source illuminate both the
accretion disc and the collecting area of a distant observer, 
whose distance from the source is taken to be $r_{\rm{obs}} = 10^3 r_g$ for 
numerical purposes. The observer measures both the direct flux 
and the reflected emission from the disc. The observer
inclination has been fixed to $30$ degrees, which is appropriate for
the case of MCG--6-30-15. 

In this work, we assume that the direct continuum emission from the
primary source dominates over the reflected continuum emission, which
is neglected in our computations. However, since the reflected
continuum varies together with the iron line, it does not affect the
results on the variability of the PLC and of the RDC (hereafter
represented by the iron line only), which are the main purpose of this
Letter.

Each simulation is carried out by sending a given number $N$ of photons
from the primary source and by integrating along the null geodesics in
the Kerr spacetime until the photons reach the accretion disc (or are
lost into the black hole horizon) or the observer collecting area.
This procedure allows us to compute directly the PLC at the observer
and the illuminating flux on the disc as a function of the photons 
incident angle $\theta_i$ and energy $E_d$, taking into account both 
special and general relativistic effects.

The iron line local intensity is computed by assuming that the
fluorescent photons are produced by cold, non--ionised matter at the
rest frame energy of $6.4$~keV and by making use of the work by George
\& Fabian (1991) who provided analytical approximations for the
dependence of the fluorescent emission on $\theta_i$ and $E_d$ (see
also Ruszkowski 2000; Lu \& Yu 2001 for further details). In this work, the iron line
emission is assumed to be isotropic in the rest frame of the disc,
which is not a strong limitation for low inclination objects such as
MCG--6-30-15.

To calculate line profiles and fluxes, we use a ray--tracing technique
to follow the trajectories of photons from the observer until they
either intersect the accretion disc or the black hole horizon and we
calculate the redshift factor corresponding to each arrival
position on the disc \cite{l91,detal97}. Then the line flux is
obtained by making use of the relativistic invariance of $I_\nu/\nu^3$,
where $I_\nu$ is the specific line intensity, and by integrating over the
solid angle subtended by the whole disc at the observer.  

\section{PLC and RDC variability}

%%%%%%%%%%%% 
\begin{figure}
\psfig{figure=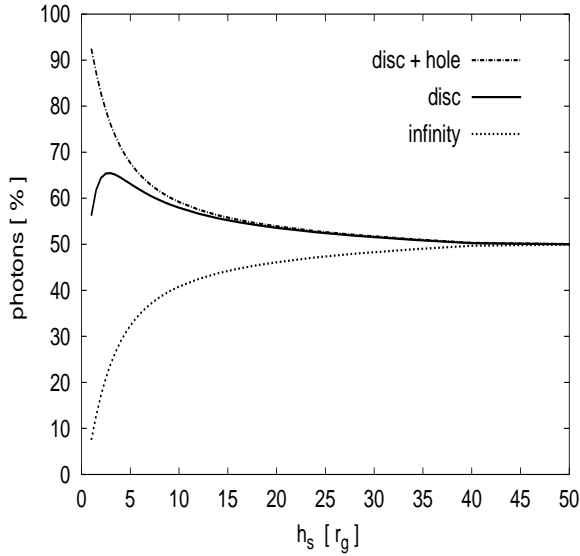,width=8.0cm,height=7.5cm,angle=-90}    
\caption{The percentage of the photons emitted by the primary
  source that reach the accretion disc or that are able to escape at
  infinity is plotted as a function of the source height above the
  equatorial plane. As a reference, the photons that either reach the disc or
  are lost into the black hole horizon is also shown. The source is
  static and it is located at a distance $\rho_s = 2.0~r_g$ from the
  rotation axis. The outer radius of the accretion disc 
  is, in this case, chosen to be $r_{out} = r_{obs}$.
}
\label{fig1}
\end{figure}

We investigate the effects that the position and the
motion of the primary source have on the continuum and reflected 
observed fluxes, with particular emphasis on the associated variability of 
both the PLC and RDC components. 

As already pointed out in previous studies 
\cite{mm96,mkm00,mmk02,dl01}, if the primary source is very close to
the central massive black hole, a large fraction of the emitted
photons will be bent onto the disc (or lost into the hole event
horizon), reducing the observed direct flux with respect to the 
reflected one and enhancing the iron line equivalent width (EW). 

This effect is shown in Fig. \ref{fig1} where we consider the
arrival positions of the photons emitted from a static source located
at $\rho_s = 2.0~r_g$ as a function of the height ($h_s$) of the source. As an
example, for a source height of $h_s = 2~r_g$, the percentage of
photons that reach the accretion disc is 64~per cent. The remaining 36
per cent is equally distributed between photons that are lost in the
hole horizon and those that reach the observer at infinity. Similar
results are obtained also in the case of a corotating source. However
in the latter case, fewer photons are lost into the hole horizon (8
per cent) due to the strong beaming in the direction of the source motion
that reduces the number of photons directed towards the hole. The
illumination pattern on the disc is then more extended for a
corotating source (74 per cent of the photons hit the disc) than for a
static one.

Here, the PLC variability is supposed to be caused by
variations of the primary source position. Since the source
height controls the amount of direct radiation that reaches 
the observer (see Fig. \ref{fig1}), the observed PLC
from a source close to the black hole is much smaller than 
that of a source located at larger height. At the same time, the RDC
responds to a much larger region than the PLC, especially if the
source is corotating, and it is likely to vary
less than the power law component. 

%%%%%%%%%%%%%
\begin{figure}
\psfig{figure=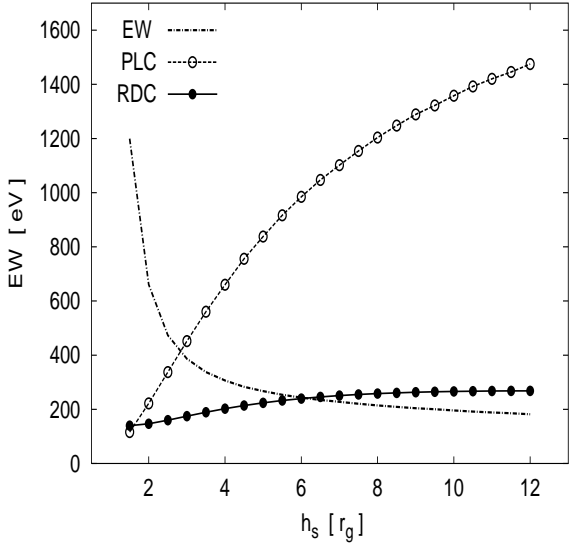,width=8.0cm,height=7.5cm,angle=-90}    
\caption{PLC and RDC components as a function of the source
  height. The iron line EW $\equiv$ RDC/PLC is also shown. 
  The source is static and it is located at $\rho_s = 2.0~r_g$
  from the rotation axis. Units of flux are arbitrary while the EW is
  given in eV. }
\label{fig2}
\end{figure}
%%%%%%%%%%%%%%
\begin{figure}
\psfig{figure=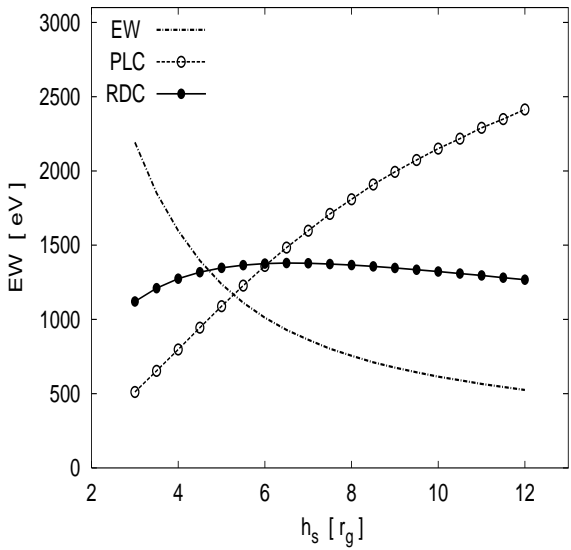,width=8.0cm,height=7.5cm,angle=-90}    
\caption{Same as Fig. \ref{fig2} but for a corotating source at 
$\rho_s = 2.0~r_g$.
}
\label{fig3}
\end{figure}
%%%%%%%%%%%%%%

To illustrate this
behaviour, and to compare the variation of the two components we
present results for sources located at $\rho_s = 2~r_g$ from the
rotation axis, but similar results can be obtained within 
$\rho_s \simeq 3-4~r_g$. 
In Figs. \ref{fig2} and \ref{fig3} we show the PLC and RDC
fluxes as a function of the source height for a static and a
corotating source respectively. The PLC flux is defined as 
the observed flux of the power law component evaluated at 
$6.4$~keV, while the RDC is the total iron
line observed flux (i.e. the integral over energy of the line). 
We also show the iron line equivalent width  defined as the ratio
between the RDC and the PLC. The general behaviour is that the
variation in EW is dominated by the PLC variability and  is
clearly anti--correlated with flux. 
Since we neglect the contribution of the reflected continuum, 
the values of the line EW are overestimated,
especially in the most extreme cases (low source height) and they   
have to be taken only as an indication. 

Both the power law and the reflection--dominated components vary with
the height of the primary source, but the variation of the RDC has a
much smaller amplitude. This behaviour is enhanced in the corotating
case shown in Fig. \ref{fig3} where, if the source height changes
from $3$ to $8~r_g$, a variation of the PLC by 
almost a factor $4$ is accompanied by a variation in the RDC with
a maximum amplitude of 15 per cent around its mean value. Even
larger variations of the PLC are found for smaller values of $\rho_s$,
together with an almost constant RDC. These results can be compared
with the work by Fabian \& Vaughan (2003) who analysed the spectral
variability of MCG--6-30-15 finding variations of the RDC up to
about 25 per cent within a PLC change of about a factor $4$. 

We note here that the two component model also accounts (Vaughan et al
2003, in preparation) for the shape and amplitude of the
r.m.s. variability spectrum of MCG--6-30-15, plotted in Fabian et al. (2002).

\subsection{Emissivity and line profiles}

%%%%%%%%%%%%%%
\begin{figure}
\psfig{figure=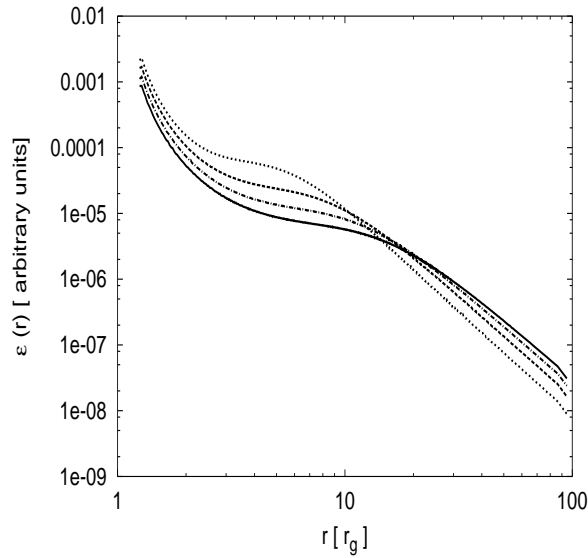,width=8.0cm,height=7.5cm,angle=-90}    
\caption{The local iron line emissivity profile on the accretion disc
  for a corotating source at $\rho_s = 2.0~r_g$ is shown for different
  values of the source height. Looking at the right side of the plot
  the source heights are, from bottom to top, $h_s = 4 ,6,8,10$. 
}
\label{fig4}
\end{figure}
%%%%%%%%%%%%%%

We compute the local iron line emissivity profile induced by a ring--like
primary source by rotating the illumination pattern of the one for a 
point--like source over the azimuthal position on the disc. The
emissivity is thus axisymmetric and it is a function of the radial
position on the disc only. The emissivity $\epsilon (r)$ produced by a
corotating ring source at $\rho_s = 2~r_g$ is shown in Fig.
\ref{fig4} for four different source heights. 
Due to the anisotropy of incident radiation (controlled by $h_s$), 
the emissivity in the inner regions is reduced by increasing the
source height, while it increases in the outer disc (see also
Martocchia, Karas \& Matt 2000). 

At a fixed $h_s$, the emissivity is
steeper in the inner disc than in the outer regions with a transition
region between about $3$ and $10~r_g$. This is in good agreement with
a recent analysis \cite{fetal02} where a best fit to the iron line of 
MCG--6-30-15 was found by using a broken power law emissivity with a
break radius of about $6~r_g$. The outer emissivity index was found to 
be $\sim 2.5$, while a steeper index of about $\sim 4.8$ was derived 
within $6~r_g$.      

%%%%%%%%%%%%%%
\begin{figure}
\psfig{figure=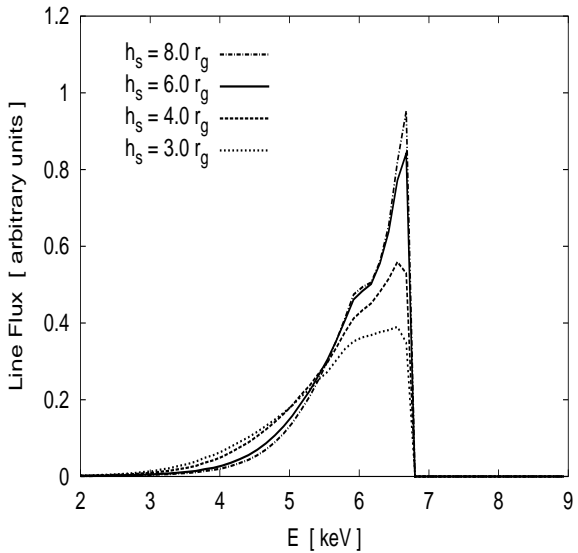,width=8.0cm,height=7.5cm,angle=-90}    
\caption{Iron line profiles for a corotating source at 
$\rho_s = 2.0~r_g$ and for different source heights. 
}
\label{fig5}
\end{figure}
%%%%%%%%%%%%%%

The iron line profiles are broad with an extended red wing if the
primary source is located close to the black hole, while the profile
becomes progressively narrower as the distance from the hole is
increased. This is due to the different illumination pattern on the
accretion disc (and different emissivity) for different source
heights. If the source is low, the inner regions of the disc
are more illuminated resulting in the enhancement of the relativistic
broadening of the iron line. Our model then predicts broader 
line profiles when the source flux is low (source at low
heights). This is indeed the sense of the changes seen in the ASCA
1994 observation \cite{ietal96}. In Fig. \ref{fig5}  we
show the line profiles obtained in the case of a corotating source at 
$\rho_s = 2.0~r_g$ for different heights in the range $3-8~r_g$, whose
variation accounts for a change in the PLC by a factor $\sim 4$. 

The changes in the line profiles for different source
heights are subtle and it could be difficult to detect them in
real X-ray data. If this is the case, an observer would conclude that
the line is almost constant both in flux and in spectral shape while
the PLC varies in normalisation, as it seems to be the case in the
recent {\it{XMM-Newton}} observation of MCG--6-30-15 \cite{fetal02} that
was used to study the source variability \cite{fv03}. 
This possibility has been tested by comparing the line profiles we
produced with this same data set. We analysed the $3$--$10$ keV 
time--averaged spectrum obtained with the
{\small{EPIC}} pn camera using {\small{XSPEC}} version 11 \cite{a96}. 
The spectrum was fitted using a simple power law plus the relativistic line 
produced by our model when a corotating source is placed at 
$\rho_s = 2.0~r_g$ with variable height between $3$ and $8~r_g$. 

The $\chi^2$ difference between the two more extreme heights is 
$\Delta\chi^2 \simeq 13$ with  $\chi^2 = 265.8 $ for  
$h_s = 3~r_g$ and $\chi^2 = 252.9$ for $h_s = 8~r_g$, both with 
$178$ degrees of freedom. However, this simple model leaves some 
residuals around $6.4$~keV and the significance of the preferred
source height is poor. These residuals can be accounted for by an 
additional narrow Gaussian component with an equivalent width of only 
$\sim 30$~eV that can be due to reflection from distant material and is 
not included in our model for line emission from the inner $100~r_g$ of the 
accretion disc. With the addition of this component, the emission line in the
{\it{XMM-Newton}} data is well fitted by our model and, most
remarkably, the $\chi^2$ for  the most extreme heights 
is now  identical (both models have 
$\chi^2 / \nu = 190.5 / 175 \simeq 1.09$). 

This means that, once the narrow line component is included, 
the changes in the line shape are substantially reduced and may be 
difficult to detect with current instruments. If the narrow component
is emitted from distant material, it is likely to remain
constant. Future observations of the source may be capable of 
resolving the narrow component  from the broader one 
emitted from the accretion disc, allowing more stringent
constraints to be placed on the height of the illuminating primary source. 
In summary, the model we have explored consistently accounts for the 
uncorrelated variability of the PLC and RDC components 
and produces also iron line profiles in good
agreement with the most recent X--ray observation of 
MCG--6-30-15.  

\section{Conclusions}

The spectral variability of MCG--6-30-15 can be explained by an
almost constant Reflection--Dominated Component together with a
highly variable Power Law Component. 

Here we have explored a model in which the variability of the PLC is
accounted for by changes in the height of the primary source above the
accretion disc. Light bending by the gravitational field of the black
hole substantially reduces the PLC flux at low heights, allowing for high
continuum variability. The iron line flux, which represents the RDC,
is much more constant with an amplitude of variation of about
15--20 per cent when the PLC varies by a factor $\sim 4$. 

It should be noted that the region where the  flare (or multiple
flares) creating the PLC should be located to reproduce this behaviour
is close to the rotation axis, within $3$-$4$ gravitational radii. It
is most plausibly powered by magnetic fields from the disc, possibly
linking to the hole (e.g. Blandford \& Znajek 1977; Wilms et al 2001), and
perhaps forming the base of a jet (note that the radio emission from
MCG--6-30-15 is weak and unresolved, Morganti et al. 1996). 

We also find that the iron line emissivity on the accretion 
disc is consistent with a broken power law profile with a transition
between about $3$ and $10~r_g$, in good agreement with recent
observations. 

In summary, the strong light bending close to the black hole can cause
the PLC and RDC to appear disconnected. Even if the source of the PLC
is intrinsically constant in luminosity, it will appear to an outside
observer to vary if its height from the disc
changes. Much of the radiation is bent onto the disc in a manner which
means that the flux of the RDC will appear constant. Subtle profile
changes are expected in the sense that the line will appear
narrower when the PLC is bright and broader when dim, but they 
are difficult to detect with current instruments.   

It is of course possible that the source of the PLC undergoes
intrinsic luminosity variations on both short and long timescales.
These will lead to connected variability between the PLC and RDC, and
reverberation on short timescales. For MCG--6-30-15 these timescales
are probably less than $10^3$~s, too short to be investigated with
XMM-Newton but within the grasp of Constellation-X.

\section*{Acknowledgements}
We would like to thank the referee, Dr. Giorgio Matt, for his helpful 
comments. Some of the results have been derived using the HEALPix package 
\cite{healpix}. ACF thanks the Royal Society for support.

\bsp

\label{lastpage}

\end{document}